\documentclass[twocolumn,showpacs,preprintnumbers,amsmath,amssymb]{revtex4}
\usepackage{amsfonts}% Physical Review B

\usepackage{enumitem}
\usepackage{epsf}
\usepackage{anysize}
\marginsize{1 in}{1 in}{1in}{1.40in}

\usepackage{graphicx}% Include figure files
\usepackage{subfigure}
\usepackage{dcolumn}% Align table columns on decimal point
\usepackage{bm}% bold math
\usepackage{amsmath}
\usepackage{amssymb}
\usepackage{mathrsfs}
\usepackage[latin1]{inputenc}
\usepackage{pstricks,pst-node,pst-text,pst-3d}
\usepackage[all]{xy}
\usepackage{diagbox}
\usepackage[ ruled ]{algorithm2e}
\usepackage{algpseudocode}

\makeatletter
  \newcommand\figcaption{\def\@captype{figure}\caption}
  \newcommand\tabcaption{\def\@captype{table}\caption}
\makeatother

\begin{document}

\title{Robust Quantum Control in Games: an Adversarial Learning Approach}

\author{Xiaozhen Ge}
%\author{Xiaozhen Ge$^{1}$, Haijin Ding$^{1}$, Herschel Rabitz$^{2,*}$, and Rebing Wu$^{1,3,\dagger}$}
\affiliation{Department of Automation, Tsinghua University, Beijing 100084, China}

\author{Haijin Ding}
\affiliation{Department of Automation, Tsinghua University, Beijing 100084, China}

\author{Herschel Rabitz}
\email{hrabitz@princeton.edu}
\affiliation{Department of Chemistry, Princeton University, Princeton, New Jersey 08544, USA}

\author{Rebing Wu}
\email{rbwu@tsinghua.edu.cn}
\affiliation{Department of Automation, Tsinghua University, Beijing 100084, China}
\affiliation{Beijing National Research Center for Information Science and Technology, Beijing 100084, China}

\begin{abstract}
	
High-precision operation of quantum computing systems must be robust to uncertainties and noises in the quantum hardware. In this paper, we show that through a game played between the uncertainties (or noises) and the controls, adversarial uncertainty samples can be generated to find highly robust controls through the search for Nash equilibria (NE). We propose a broad family of adversarial learning algorithms, namely a-GRAPE algorithms, which include two effective learning schemes referred to as the best-response approach and the better-response approach within the game-theoretic terminology, providing options for rapidly learning robust controls. Numerical experiments demonstrate that the balance between fidelity and robustness depends on the details of the chosen adversarial learning algorithm, which can effectively lead to a significant enhancement of control robustness while attaining high fidelity.
\end{abstract}

\maketitle

\section{introduction}
 Recent experimental breakthroughs in quantum computing have signaled its commercialization in the foreseeable future~\cite{Castelvecchi2017Quantum, Dai2017Quantum, Mohseni2017Commercialize}. The demands involved require highly precise and stable control techniques for deterministic implementation of quantum gates~\cite{nielsen2002quantum}. Generally, high-precision is relatively easy to achieve if a well characterized model is available, for example, using the highly efficient GRAPE (gradient ascent pulse engineering) algorithm~\cite{Khaneja2005Optimal} and methodologies based on reinforcement learning~\cite{dalgaard2019global}. The real challenge is to maintain high precision in the presence of realistic uncertainties and noises in the model, i.e., finding both high-precision and robust controls. In the literature, a variety of proposals have been put forth for either online or offline searches for robust quantum controls, e.g., stimulated Raman adiabatic passage working to overcome pulse shape errors~\cite{bergmann1998coherent, bergmann2015perspective,vitanov2017stimulated}, dynamical decoupling to fight against decoherence noises~\cite{Viola1999Dynamical,khodjasteh2009dynamically, khodjasteh2007performance, Khodjasteh2005Fault}, and differential evolution~\cite{Zahedinejad2016Designing,Ehsan2015High} as well as ensemble-based algorithms~\cite{Khaneja2005Optimal,Li2006Control} aiming to deal with a variety of noises and uncertainties .

Recently, a class of robust control design methods have been proposed~\cite{Zhang2014Robust,Chen2014Sampling,Dong2013Sampling,Wu2017Robust,Wu2018Deep}. %.based on gradient algorithms using 
It was shown that, by Monte Carlo sampling of the uncertainties or noises, the GRAPE algorithm can be exploited to effectively improve the robustness against linewidth broadening~\cite{Chen2014Sampling,Zhang2014Robust,Dong2013Sampling}, amplitude errors of control fields~\cite{Chen2014Sampling,Dong2013Sampling}, coupling uncertainty~\cite{Wu2018Deep}, and clock noises~\cite{ding2019robust}.

 Most of the above approaches are based on improving robustness of a quantum control protocol quantified by the average performance with respect to the system uncertainties or noises to be suppressed. This measure is relatively easy to evaluate and hence to optimize. However, it is not the unique choice, nor necessarily the best. For example, fighting against the worst-case performance was widely adopted in the control of classical systems~\cite{stoorvogel1993robust,Yoon2005On,Shi2013Gain,Poznyak2002}, which leads to a class of min-max problems. The optimization with respect to such an objective can effectively reduce the risk of failure, which is demanded by fault-tolerant quantum computation. The worst-case optimization for robust quantum controls was first formulated in the control of molecular systems against disturbance~\cite{Beumee1992Robust,zhang1994robust}, and a theoretical analysis was later done on the robust performance via the Dyson expansion \cite{Koswara2014}. However, there are very few algorithms for efficiently solving the associated min-max problem. In \cite{James2007H,Petersen,Maalouf2017Finite}, the so-called $H_{\infty}$ approach was applied to a class of linear quantum systems in the Heisenberg picture. In \cite{kosut2013robust}, sequential convex programming (SCP) was proposed for solving the worst-case robust optimization problem in a single-qubit system, which decomposes the min-max problem into a sequence of convex optimization procedures.  

Since the min-max problem can be taken as an adversarial game between the control and the uncertainties (or noises) that attempt to affect the objective of the quantum system (e.g., state or gate infidelity) in opposite directions, one can introduce game-theoretic learning algorithms that seek Nash equilibria (NE)~\cite{von2007theory}. Even if the Nash equilibrium does not exist or is hard to be reached, the robustness of the control can still be enhanced during the learning process. As the adversarial game has a close relationship with both the fidelity and robustness, the game-based learning process can be adjusted flexibly to make a balance between the two performance indexes, forming a broad family of adversarial learning algorithms. A successful paradigm is the GAN (Generative Adversarial Nets) model in deep learning~\cite{szegedy2013intriguing,understanding,explaining,generative} for generative learning tasks, where a large number of applications show that the learning model can be trained to be more {\it generalizable} by actively generating adversarial samples produced by a discriminator neural network.

In this paper, we will extend game-theoretic ideas to the design of robust quantum controls, starting with the worst-case system disturbance. The remainder of this paper will be arranged as follows. Sec.~\ref{analysis} presents a game-theoretic analysis of the robust control problem.
In Sec.~\ref{algorithm}, we introduce several adversarial learning  control algorithms based on the dynamic processes in the game theory of learning and evolution. In Sec.~\ref{experiment}, the effectiveness of the proposed algorithms is illustrated through simulations of the control design in two representive examples. Finally, concluding remarks are made in Sec.~\ref{conclusion}.

\section{The Zero-sum Game between control and uncertainty}\label{analysis}

Consider an $N$-dimensional quantum control system whose unitary propagator obeys the following Schr\"{o}dinger equation:
\begin{equation}
      \dot{U}(t;\boldsymbol{u},\boldsymbol{\epsilon}) = -iH\left[t;\boldsymbol{u},\boldsymbol{\epsilon}\right]U(t;\boldsymbol{u},\boldsymbol{\epsilon})
\end{equation}
where $U(\cdot)\in \mathbb{C}^{N\times N}$ represents the quantum gate operation with the initial value being the identity matrix. The system's Hamiltonian $H\left[t;\boldsymbol{u},\boldsymbol{ \epsilon}\right]$ %(assumed to be traceless for all $t$, ${\bf u}$ and ${\bf \epsilon}$)
is dependent on a vector of control parameters $\boldsymbol{u}$ (e.g., in-phase and quadrature amplitudes, or phases and amplitudes of laser pulses in the frequency-domain) and a vector of uncertainty parameters $\boldsymbol{\epsilon}$ (e.g., environmental noises or imprecisely identified parameters). %that obeys some probability density distribution $P(\epsilon)$.

Let $U_f$ be the target gate operation and
\begin{equation}\label{}
 L[\boldsymbol{u},\boldsymbol{\epsilon}]=N^{-2}\|U(T;\boldsymbol{u},\boldsymbol{\epsilon})-U_f\|^2
\end{equation}
be the infidelity of the controlled gate under the control $\boldsymbol{u}$ and the uncertainty $\boldsymbol{\epsilon}$, where $\|\cdot\|$ is the Frobenius norm. The robustness objective is to find a control $\boldsymbol{u}$ under which
$L[\boldsymbol{u},\boldsymbol{\epsilon}]$ is as small as possible for as many as possible values of the uncertainty ${\boldsymbol{\epsilon}}$.

A straightforward approach adopted in most existing studies is to minimize the average infidelity, i.e.,
\begin{equation}\label{average error}
{\bf E}_{\boldsymbol{\epsilon}}L[\boldsymbol{u},\boldsymbol{\epsilon}]=\int_{\boldsymbol{\epsilon}}L[\boldsymbol{u},\boldsymbol{\epsilon}]p_0(\boldsymbol{\epsilon}){\rm d}\boldsymbol{\epsilon},%\int L[\boldsymbol{u},\boldsymbol{\epsilon}]P(\boldsymbol{\epsilon}){\rm d}{\bf \epsilon},
\end{equation}
where $p_0(\boldsymbol{\epsilon})$ is the probability density distribution of $\boldsymbol{\epsilon}$. This objective is relatively easy to estimate and thus to optimize, but the resulting control may not be able to dictate all possible cases of uncertainties due to a  lack of control over the variance of the infidelity (i.e., the infidelity can be high over certain small ranges of uncertainties even if the average infidelity is low).

To reduce the risk of encountering high infidelity, we consider the worst-case performance instead of the average performance, which leads to the following min-max problem:
\begin{equation}\label{problem1}
\min_{\boldsymbol{u}}\max_{\boldsymbol{\epsilon}}L[\boldsymbol{u},\boldsymbol{\epsilon}]
\end{equation}
that does not rely on the probability density distribution of $\boldsymbol{\epsilon}$. Once the worst-case infidelity is below the desired threshold value, the risk of high infidelity can be effectively reduced.

From a game-theoretic point of view, the optimization process can be taken as a zero-sum game between two players, the control $\boldsymbol{u}$ and the uncertainty $\boldsymbol{\epsilon}$, in which $\boldsymbol{u}$ attempts to reduce the gate infidelity while $\boldsymbol{\epsilon}$ tries to increase it. A robust control is thus naturally associated with the NE point $(\boldsymbol{u}^*, \boldsymbol{\epsilon}^*)$ of the game~\cite{von2007theory}, at which each player is unable to go any further by tuning merely $\boldsymbol{u}$ or merely $\boldsymbol{\epsilon}$. That is to say, it holds that 
\begin{equation}\label{con}
	\boldsymbol{u}^*=\arg\min_{\boldsymbol{u}}L[\boldsymbol{u},\boldsymbol{\epsilon}^*]\ \text{and} \
	\boldsymbol{\epsilon}^*=\arg\max_{\boldsymbol{\epsilon}}L[\boldsymbol{u}^*,\boldsymbol{\epsilon}].
	\end{equation}
  A standard approach to search for such a strategy profile is to alternately minimize (with respect to $\boldsymbol{u}$) and maximize (with respect to $\boldsymbol{\epsilon}$) the infidelity. %with respect to $\boldsymbol{u}$ and $\boldsymbol{\epsilon}$, respectively. 
Even if the NE does not exist or is hard to be reached, the robustness of the control may still be enhanced during the process of fighting against the so-called {\it adversarial} samples of $\boldsymbol{\epsilon}$ that yield the worst performance. %The maximization process generates the so-called {\it adversarial} samples of $\boldsymbol{\epsilon}$ that yield the worst performance, against which the control robustness can be enhanced via the minimization processes. 
In this spirit, a family of learning algorithms can be devised in which the uncertainty parameters play a more active role instead of just being averaged out. Since in this work the gradient-based GRAPE algorithm is always applied in the minimization process, the adversarial learning algorithms to be presented below will be termed as a-GRAPE, where ``a" stands for ``adversarial".

%It should be noted that the order of actions taken by the two players is not specified in (\ref{con}), which implies that the solution of the min-max problem (\ref{problem1}) with ordered action must satisfy the condition (\ref{con}), but the profile satisfying (\ref{con}), may not be a solution of (\ref{problem1}). Nevertheless, the NE-gradient optimization can still effectively improve the robustness of controls, and it does not matter whether the solution of (\ref{problem1}) is reached or not.

\section{The Design of Adversarial Learning control algorithms }\label{algorithm}
In this section, we will propose two types of a-GRAPE algorithms, namely the best-response and better-response approaches, for training highly robust controls via active selection of adversarial samples.

\subsection{Best-response approach}

The simplest NE-seeking approach consists of rounds of alternate optimization with the control and the uncertainty. Suppose that we have obtained an adversarial sample $\boldsymbol{ \epsilon}^{(k)}$
subject to the optimal control $\boldsymbol{u}^{(k)}$ in the $k$-th round. In the $(k+1)$-th round, we first minimize $L[\boldsymbol{u},\boldsymbol{\epsilon}^{(k)}]$ with respect to $\boldsymbol{u}$ using the GRAPE algorithm, which updates the control by $\boldsymbol{u}^{(k+1)}$. Then, we update the adversarial sample of uncertainty parameters by $\boldsymbol{\epsilon}^{(k+1)}$ that maximizes $L[\boldsymbol{u}^{(k+1)},\boldsymbol{\epsilon}]$ with respect to $\boldsymbol{\epsilon}$.

Utilizing game theory terminology, we call such an adversarial learning process a best-response approach because each player chooses its best strategy against the opponent~\cite{fudenberg1998theory}. %This learning strategy had been very successful in the training of generative adversarial networks (GANs) for learning the distribution of data~\cite{generative}. 
However, for most robust quantum control problems, the best-response strategy can hardly reach a pure NE. This conclusion follows from the facts that a NE requires $L[\boldsymbol{u}^*,\boldsymbol{\epsilon}]=0$ for all admissible $\boldsymbol{\epsilon}$, as the control is assumed to have fully adequate resources (e.g., bandwidth) such that $\min_{\boldsymbol{u}}L[\boldsymbol{u},\boldsymbol{\epsilon}]=0$ for any $\boldsymbol{\epsilon}$, and that this condition is hard to satisfy in practice unless $\boldsymbol{\epsilon}$ is only allowed to vary over a very small domain. Since a min-max problem possesses a NE when the minimization problem is convex and maximization problem is concave, a viable strategy is to seek a mixed NE in a enlarged domain, where the uncertainty $\boldsymbol{\epsilon}$ is allowed to adopt mixed strategies, i.e., instead of picking a single adversarial sample, we generate a distribution of adversarial samples. Mathematically, this leads to the following min-max problem:
\begin{equation}\label{problem2}
\min_{\boldsymbol{u}}\max_{p(\boldsymbol{\epsilon})\in \mathcal{P}}\int L[\boldsymbol{u},\boldsymbol{\epsilon}]p(\boldsymbol{\epsilon}){\rm d}\boldsymbol{\epsilon},
\end{equation}
where the maximization is performed over the space of probability density distributions $\mathcal{P}$. The mixed NE $(\boldsymbol{u}^*, p^*(\boldsymbol{\epsilon}))$ satisfies
{\setlength\abovedisplayskip{1pt}
	\setlength\belowdisplayskip{1pt}
\begin{eqnarray*}
\boldsymbol{u}^*&=&\arg\min_{\boldsymbol{u}}\int L[\boldsymbol{u},\boldsymbol{\epsilon}]p^*(\boldsymbol{\epsilon}){\rm d}\boldsymbol{\epsilon},\\
p^*(\boldsymbol{\epsilon})&=&\arg\max_{p(\boldsymbol{\epsilon})}\int L[\boldsymbol{u}^*,\boldsymbol{\epsilon}]p(\boldsymbol{\epsilon}){\rm d}\boldsymbol{\epsilon}.
\end{eqnarray*}}
The advantage of searching for a mixed NE is that their existence is more likely ensured by the linear dependence in $p(\boldsymbol{\epsilon})$ and hence a mixed NE is easier to find.

However, problem (\ref{problem2}) is computationally much more expensive than problem (\ref{problem1}) because the search space for the maximization part is much larger. In practice, this issue can be relaxed by approximating the optimal probability distribution in order to simplify the maximization process.

Here, we propose that the optimal distribution  $p(\boldsymbol{\epsilon})$ can be approximated by exploiting adversarial samples found in the past rounds. In particular, we perform the maximization process in the same way as the above best-response approach and approximate the optimal probability distribution in the $(k+1)$-th round as
\begin{equation}\label{distribution}
  p^{(k+1)}(\boldsymbol{\epsilon})\approx \frac{1}{k}\sum_{j=1}^k \delta(\boldsymbol{\epsilon}-\boldsymbol{\epsilon}^{(j)})
\end{equation}
using the historic adversarial samples $\boldsymbol{\epsilon}^{(1)}, \boldsymbol{\epsilon}^{(2)}, \cdots, \boldsymbol{\epsilon}^{(k)}$ in the past $k$ rounds. Consequently, in the following minimization process, the objective function based on the distribution (\ref{distribution}) can be written as the average infidelity over the $k$ adversarial samples
\begin{equation}\label{average}
J[\boldsymbol{u},B_k]=\frac{1}{k}\sum_{j=1}^{k}L[\boldsymbol{u},\boldsymbol{\epsilon}^{(j)}],
\end{equation}
where $B_k=\{\boldsymbol{\epsilon}^{(1)},\cdots,\boldsymbol{\epsilon}^{(k)}\}.$

In practice, the approximation (\ref{distribution}) can be chosen more flexibly. For example, one does not have to use all historic adversarial samples in (\ref{average}), because it will be too costly when $k$ is large and the early samples are likely less adversarial. Thus, we keep only the latest few samples, i.e., let the algorithm utilize only a finite number, say $s$, of adversarial samples (see Algorithm~\ref{algorithm1} for a summary). In this scenario, the originally discussed best-response approach can be taken as a special case with memory size being $s=1$.

\begin{algorithm}[htb]\label{algorithm1}
	
\caption { \textit{best-response} a-GRAPE}
\noindent{\bfseries Initialize:}

  \setlength\parindent{1em} a randomly chosen initial control $\boldsymbol{u}^{(0)}$;	

  \setlength\parindent{1em} an initial uncertainty sample set $B_0=\{0\}$;
  
   \setlength\parindent{1em} a set memory size $s$.

\noindent{\bfseries Repeat:}
\begin{itemize}[itemsep=1pt, topsep=2pt,partopsep=2pt,parsep=1pt]
\item[(1)] Use the GRAPE algorithm to update the control by the optimal solution of minimizing $J[\boldsymbol{u},B_{k-1}]$, i.e.,
{\setlength\abovedisplayskip{1pt}
	\setlength\belowdisplayskip{1pt}
	\begin{equation*}
\boldsymbol{u}^{(k)}=\arg\min_{\boldsymbol{u}}J[\boldsymbol{u},B_{k-1}],
\end{equation*}
in which $\boldsymbol{u}^{(k-1)}$ is taken as the initial guess for the GRAPE algorithm. Here, $k$ is an index to the current number of round;}
\item[(2)] Generate a new adversarial sample by 
{\setlength\abovedisplayskip{1pt}
	\setlength\belowdisplayskip{1pt}
	\begin{equation*}
\boldsymbol{\epsilon}^{(k)}=\arg\max_{\boldsymbol{ \epsilon}}L[\boldsymbol{u}^{(k)},\boldsymbol{\epsilon}];
\end{equation*}}
\item[(3)] Update:\\
{\bfseries if} $|B_k|<s$, {\bfseries then} $B_k=B_{k-1}\cup\{\boldsymbol{\epsilon}^{(k)}\}$,\\
{\bfseries else}

$B_k=\{\boldsymbol{\epsilon}^{(k)},\boldsymbol{\epsilon}^{(k-1)},\cdots,\boldsymbol{\epsilon}^{(k-s+1)}\}$;
%{\bfseries end if }
\end{itemize}

\noindent {\bfseries End} if the stoping criteria are satisfied.

\end{algorithm}

\subsection {Better-response approach}
In the above best-response approach, the minimization process is usually efficient as long as the control resources (e.g., bandwidth, pulse energy, etc.) are abundant, owing to the underlying nice control landscape topology over which almost all locally optimal controls are actually globally optimal~\cite{rabitz2005landscape,russell2016quantum}. However, the generation of adversarial samples is much harder because the maximization process is usually non-concave. Here, we relax this problem by choosing strong, but not necessarily the strongest adversarial samples for the training of robust controls. In this regard, we call this method a better-response approach.

The simplest way to search for better-response adversarial samples is to randomly choose a batch of uncertainty samples, calculate their corresponding cost, and keep the worst few members among them for the adversarial training in the next round (see Algorithm~\ref{algorithm2} for a description). The batch size of the samples should be sufficiently large so that the chosen adversarial samples have members close to the worst-case samples, but not too large to maintain computational efficiency. Naturally the better-response approach relying on random sampling usually takes more rounds of gaming, but each round can be much faster when the batch is not very large. Moreover, the randomness of sample batches in the better-response approach may bring additional benefits for the search to get away from unwanted false worst-case traps. 

\begin{algorithm}[htb]\label{algorithm2}
\caption{ \textit{better-response} a-GRAPE}

\noindent{\bfseries Initialize:}

\setlength\parindent{1em} a randomly chosen initial control $\boldsymbol{u}^{(0)}$;

\setlength\parindent{1em} a set ratio $r$, $r\in(0,1)$.	

\noindent{\bfseries Repeat:}
\begin{itemize}[itemsep=1pt, topsep=2pt,partopsep=2pt,parsep=1pt]
	\item[(1)] Randomly generate $M$ uncertainty samples, compute the corresponding infidelity and form an adversarial sample set denoted as $B_k$ by retaining the first $r M$ worst ones, where $k$ denotes the current number of round;
		\item[(2)] Use the GRAPE algorithm to update the control by
		{\setlength\abovedisplayskip{1pt}
			\setlength\belowdisplayskip{1pt}
		\begin{equation*}
			\boldsymbol{u}^{(k)}=\arg\min_{\boldsymbol{u}}J[\boldsymbol{u},B_{k}],
			\end{equation*}
		where $\boldsymbol{u}^{(k-1)}$ is taken as the initial guess for the GRAPE algorithm.}
	\end{itemize}

	\noindent {\bfseries End} if the stoping criteria are satisfied.
	\end{algorithm}

\section{Simulation results}\label{experiment}
To illustrate the above game-based adversarial learning strategies for the robust control design, we simulate two quantum gate synthesis examples in this section. As will be seen in the simulations, the best-response approach can effectively suppress the worst-case performance, but does not always lead to good performance in high precision regimes, where the better-response approach is more effective. That is to say, the best-response and the better-response approaches do not mean the best performance and the better performance respectively, and which approach performs satisfactorily depends on case-specific requirements.

\subsection{Two-qubit system}
Consider a two-qubit quantum gate control problem with the following system Hamiltonian:
\begin{eqnarray*}
H(t)&=&(1+\epsilon_0)g\sigma_{1z}\otimes\sigma_{2z}+\sum_{i=1}^2(1+\epsilon_i)\cdot\\
&&\left[u_{ix}(t)\sigma_{ix}+u_{iy}(t)\sigma_{iy}\right],
\end{eqnarray*}
where $g=10$ MHz is the identified qubit-qubit coupling strength with $\epsilon_0$ being the identification error in the coupling constant; $\epsilon_1$ and $\epsilon_2$ represent the inhomogeneity of control fields.
The dimensionless three uncertainty parameters are all assumed to be bounded by $|\epsilon_i|\leq 0.2$. In the simulation, the time duration of the control pulses is chosen as $T=300 $ ns, which is evenly divided into $100$ intervals over which the control fields are piecewise constant. The target $U_f$ is set as the controlled-NOT gate.

\begin{figure}
	\begin{center}
		\includegraphics[width=1\columnwidth]{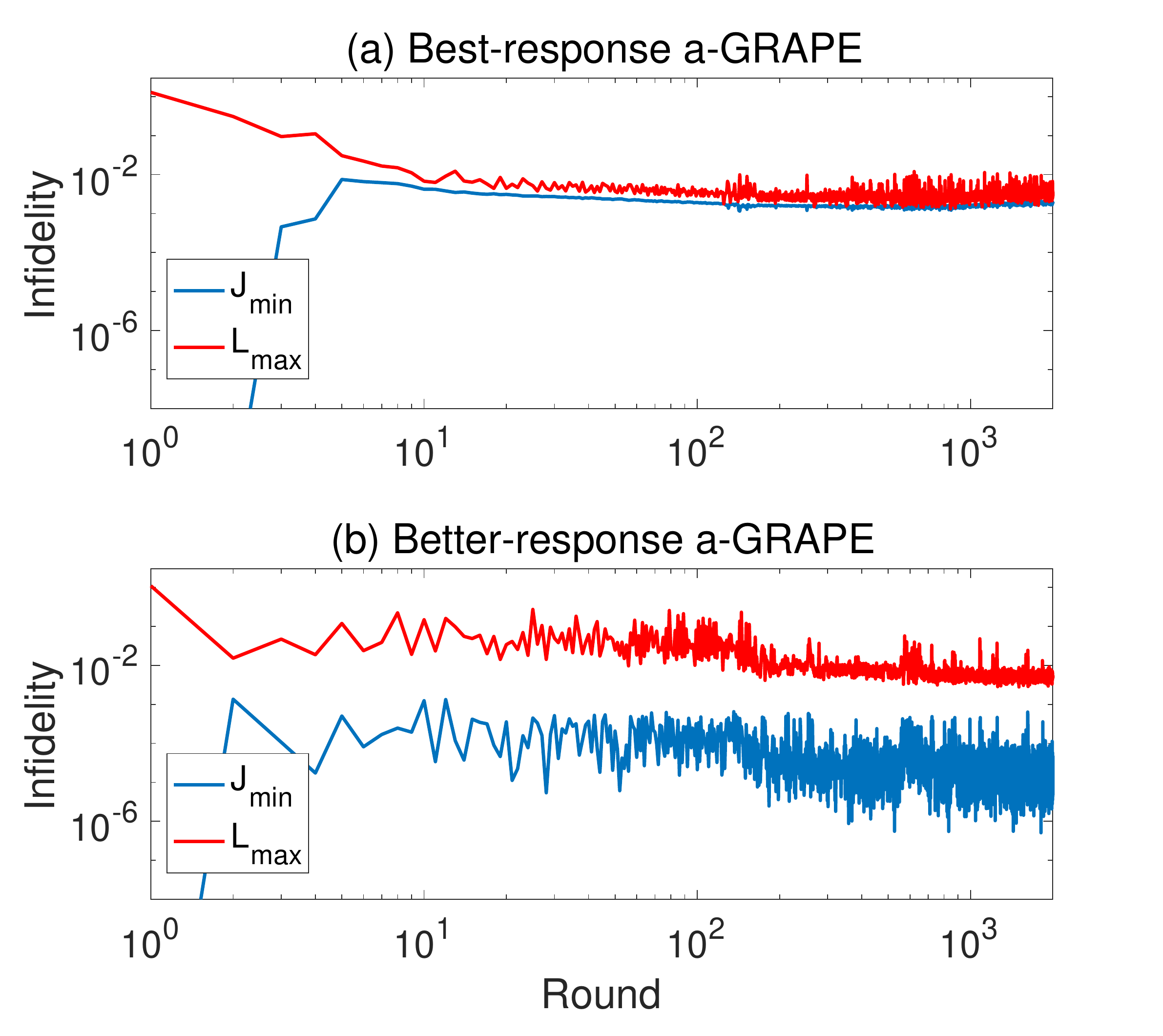}
	\end{center}
	\caption{ The learning curves of a-GRAPE for robust controls of the two-qubit system. The red curves correspond to the worst-case infidelity while the blue ones correspond to minimized average infidelity over selected adversarial samples in cases (a) the best-response approach with memory size $s=10$ and (b) the better-response approach with $M=100$ and $r=0.1$.}
	\label{learning2bit}
\end{figure}

In the best-response approach, we apply a genetic algorithm to seek adversarial samples, and use $s=10$ historic adversarial samples to train the control function in each round. In the better-response approach, we uniformly generate $M=100$ random uncertainty samples in each round and keep the first $10\%$ (i.e., $r=0.1$) worst ones as adversarial samples for training the control. Figure~\ref{learning2bit} shows the resulting learning curves, namely the achieved worst-case infidelity $L_{max}$ versus the number of rounds, as well as the corresponding minimized average infidelity $J_{min}$ over the selected adversarial samples versus the number of rounds. The robustness of the controls can be directly seen from the curves of worst-case infidelity, which are all enhanced during the optimization. The $L_{max}$-curve is initially far from the $J_{min}$-curve, but the gap is quickly reduced after several rounds of gaming. In the best-response approach, the gap is almost closed, showing that the optimized control and uncertainty samples are likely close to a mixed NE. In the better-response approach, the gap still remains large after 2000 rounds. For both approaches, the control robustness is still enhanced by the game, with the (approximate) worst-case infidelity decreased to the level of $10^{-2}$.

\begin{figure}
	\begin{center}
		\includegraphics[width=1\columnwidth]{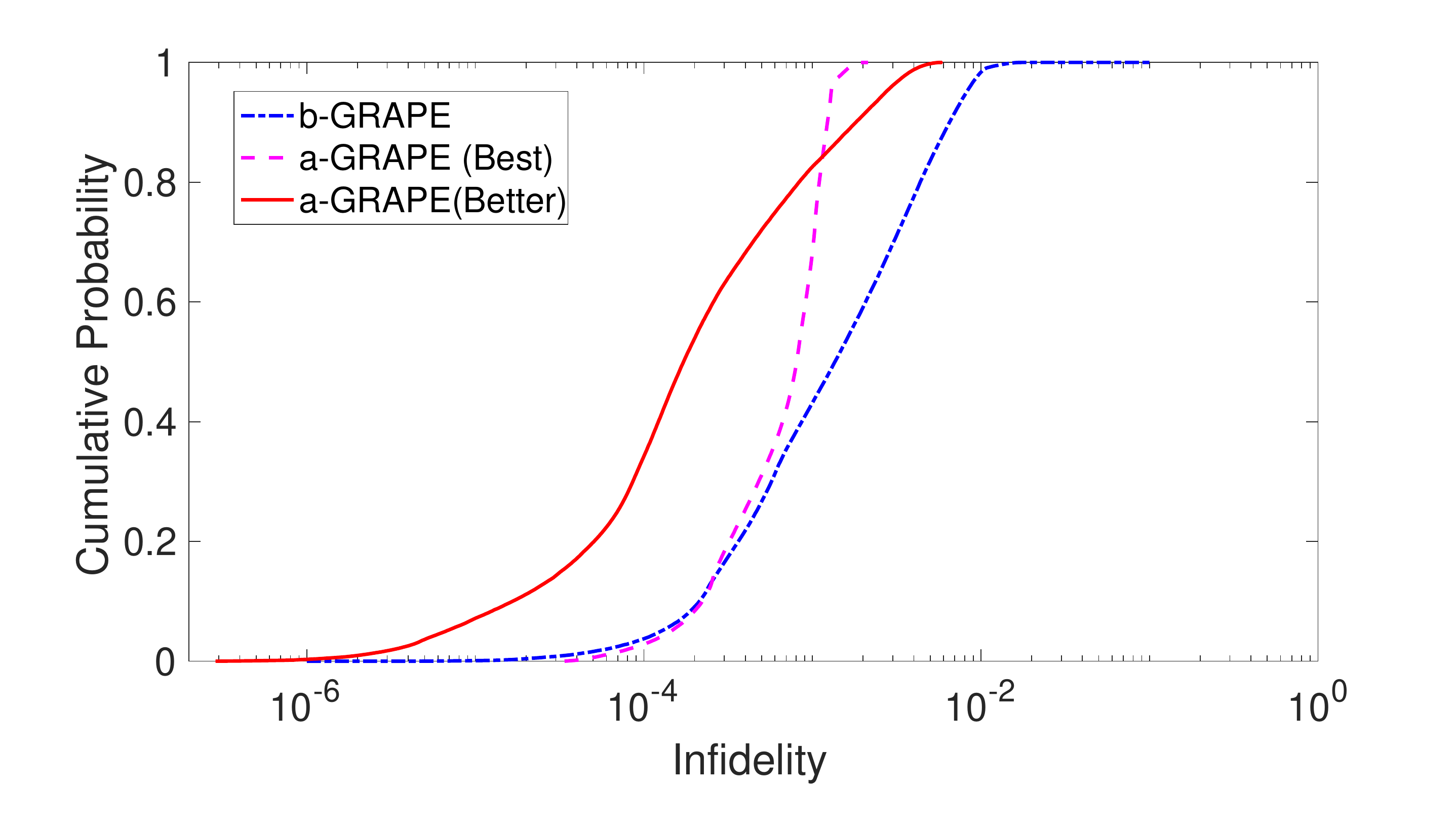}
	\end{center}
	\caption{For the two-qubit system, the cumulative probability functions $F(l)$ of the gate infidelity $l$ under the controls optimized with b-GRAPE (batch size $n_{mb}=1$), best-response a-GRAPE (memory size $s=10$), and better-response a-GRAPE ( $M=100$ and $r=0.1$).}
	\label{comparison22}
\end{figure}

As discussed above, neither the average performance nor the worst-case performance is the unique measure for quantifying the control robustness. To better evaluate the overall performance of an optimized control, we 
%To evaluate the robustness of an optimized control, we 
calculate and display the cumulative probability distribution function (cdf) $F(l)$ of the gate infidelity, i.e., the probability for the infidelity not being greater than $l$, in Fig.~\ref{comparison22}. Here, we also compare the a-GRAPE algorithms with the recently proposed b-GRAPE algorithm~\cite{Wu2018Deep} (see Appendix for details) for robust control design subject to the average infidelity (i.e., Eq.~(\ref{average error})). In the simulations, the b-GRAPE algorithm is run by $1$ million iterations having the mini-batch size $n_{mb}=1$ and a learning rate $\alpha=0.002$, while the a-GRAPE algorithms are run by 844 rounds in the best-response approach and 1504 rounds in the better-response approach.

 The cdf curve can be used to evaluate the control robustness from two perspectives. First, given a desired value of gate infidelity $l_0$ (e.g., the threshold error for quantum error correction), the cumulative probability $F(l_0)$ gives the confidence that the control can suppress the error below the value $l_0$. Second, given an expected confidence $F_0$ (say, $90\%$), the cdf can tell us at which threshold value (i.e., $l_0$ such that $F(l_0)=F_0$) the control can guarantee the confidence. As will be seen below, the robustness performance may vary at different levels of desired infidelity or expected confidence.

Figure~\ref{comparison22} clearly shows that the controls optimized by the a-GRAPE algorithms, especially by the better-response approach, are much more robust as almost the entire cdf curves are higher than that generated with the b-GRAPE algorithm (i.e., with greater confidence at each value of infidelity). For example, as seen in Table~\ref{Tab1}, the control optimized with the better-response a-GRAPE can suppress the gate error below $10^{-3}$ with a high confidence $82.5\%$, while the control optimized with b-GRAPE has only $43.1\%$ confidence. At the higher-precision level (i.e., infidelity lower than $10^{-4}$), the better-response approach still maintains $34.2\%$ confidence, while the control optimized with b-GRAPE provides only $3.7\%$ confidence. The performance of the best-response a-GRAPE is only a little poorer than b-GRAPE in the high-precision regime, but much higher in the low-precision to medium-precision regime (i.e., infidelity in $10^{-3}\sim10^{-2}$). Additionally, %consistent with the findings in Fig.~\ref{learning2bit}, 
the best-response a-GRAPE achieves a lower worst-case infidelity than that achieved by the better-response a-GRAPE, which is consistent with the finding from the $L_{max}$ curves in Fig.~\ref{learning2bit}. However, the best-response a-GRAPE has poorer performance in the high-precision regime than the better-response a-GRAPE, which is also indicated by the $J_{min}$ curves displayed in Fig.~\ref{learning2bit}.
%from which it can be also observed that the best-response approach can ensure a lower worst-case infidelity as indicated by the $L_{max}$ curves. 
The comparison between different algorithms shows that there is no unique criterion for evaluating the control robustness. An optimized control may achieve satisfactory precision (e.g., infidelity in $10^{-3}\sim 10^{-2}$) over a large regime of uncertainties, but the highest precision it can achieve may be poor. In practice, one may need a balance between the high precision and the robust regions, especially using limited control resources.

\begin{table}[!hbp]
	\centering
	\begin{tabular}{c|c|c|c}
		\hline
		\hline
		& b-GRAPE & best-response & better-response\\
		\hline
		$10^{-3}$ & $43.1\%$& $68.1\%$& $82.5\%$\\
		$10^{-4}$ &$3.7\%$ & $2.8\%$ & $34.2\%$\\
	%	\hline
	%	$99\%$ & 0.0107  & 0.0033 & 0.0041 \\
	%	$80\%$ & 0.0043 & 0.0019 &  0.0008 \\
		\hline
		\hline
	\end{tabular}
	\caption{ The two rows list the confidence for the gate infidelity to be below $10^{-3}$ and $10^{-4}$.} %The lower two rows list the gate infidelities below which $99\%$ and $80\%$ confidence can be guaranteed.}
	\label{Tab1}
\end{table}

\subsection{Three-qubit system}
To see more clearly how the control performance relies on the uncertainties and algorithmic parameters, we simulate a three-qubit system with two uncertainty parameters, whose Hamiltonian is 
\begin{eqnarray*}
H(t)&=&J_{12}(1+\epsilon_1)\sigma_{1z}\sigma_{2z}+J_{23}(1+\epsilon_2)\sigma_{2z}\sigma_{3z}\\
&&+\sum_{k=1}^3\left[u_{kx}(t)\sigma_{kx}+u_{ky}(t)\sigma_{ky}\right],
\end{eqnarray*}
where the nominal coupling constants are $J_{12}=J_{23}=10$ MHz. The uncertainty parameters $\epsilon_1$ and $\epsilon_2$ (i.e., identification errors of the coupling constants) %and $\epsilon_2$ (for field inhomogeneity) 
are bounded by $|\epsilon_{i}|\leq0.2$. In the simulation, the target unitary operation is selected as the Toffoli gate. The time period $[0,T]$, where $T=1\ \mu$s, is evenly divided into $100$ intervals, over which the control fields are piecewise constant.

\begin{figure}[!htb]
	\begin{center}
	\includegraphics[width=1\columnwidth]{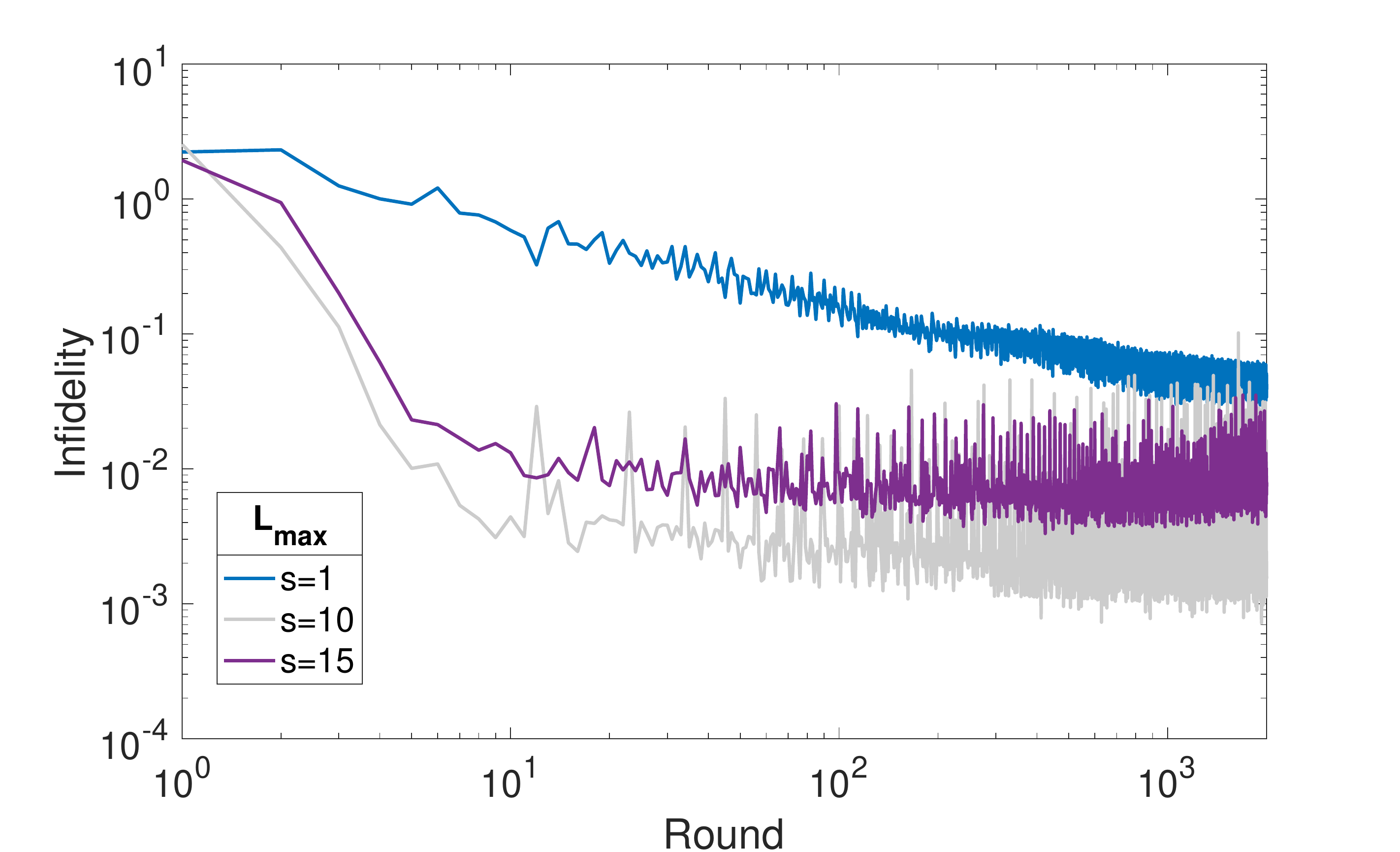}
\end{center}
\caption{ The worst-case infidelity versus the number of rounds in best-response a-GRAPE optimization of the three-qubit system with different memory sizes.
}\label{bestlearning}
\end{figure}

The previous illustration with the two-qubit system showed that the controls optimized by the a-GRAPE algorithms could improve the worst-case performance, and here we want to see how the performance depends on the parameters, e.g., $s$ and $r$. We first compare the best-response a-GRAPE optimization processes with different memory sizes $s=1,\ 10,$ and $ 15$. The learning curves are shown in Fig.~\ref{bestlearning} in which the minimization curves are not displayed because the worst-case performance is only related with the maximization curve. It can be seen that the algorithm converges faster and finds more robust controls when using more, but not too many, historic adversarial samples. For example, the worst-case infidelity reaches $10^{-2}$ after only 7 rounds when $s=10$, which converges faster than the case $s=1$, and the worst infidelity is much lower. However, the case $s=15$ performs less satisfactorily than the case $s=10$. This is reasonable because elder historic samples tend to be less adversarial due to the fading memory effect.  %Moreover, together with Fig.~\ref{bestlearning}(b), we can conjecture that using too much historical samples would also decrease the confidence of high fidelity (e.g., 0.999). 

\begin{figure}
 	\begin{center}
		\includegraphics[width=1\columnwidth]{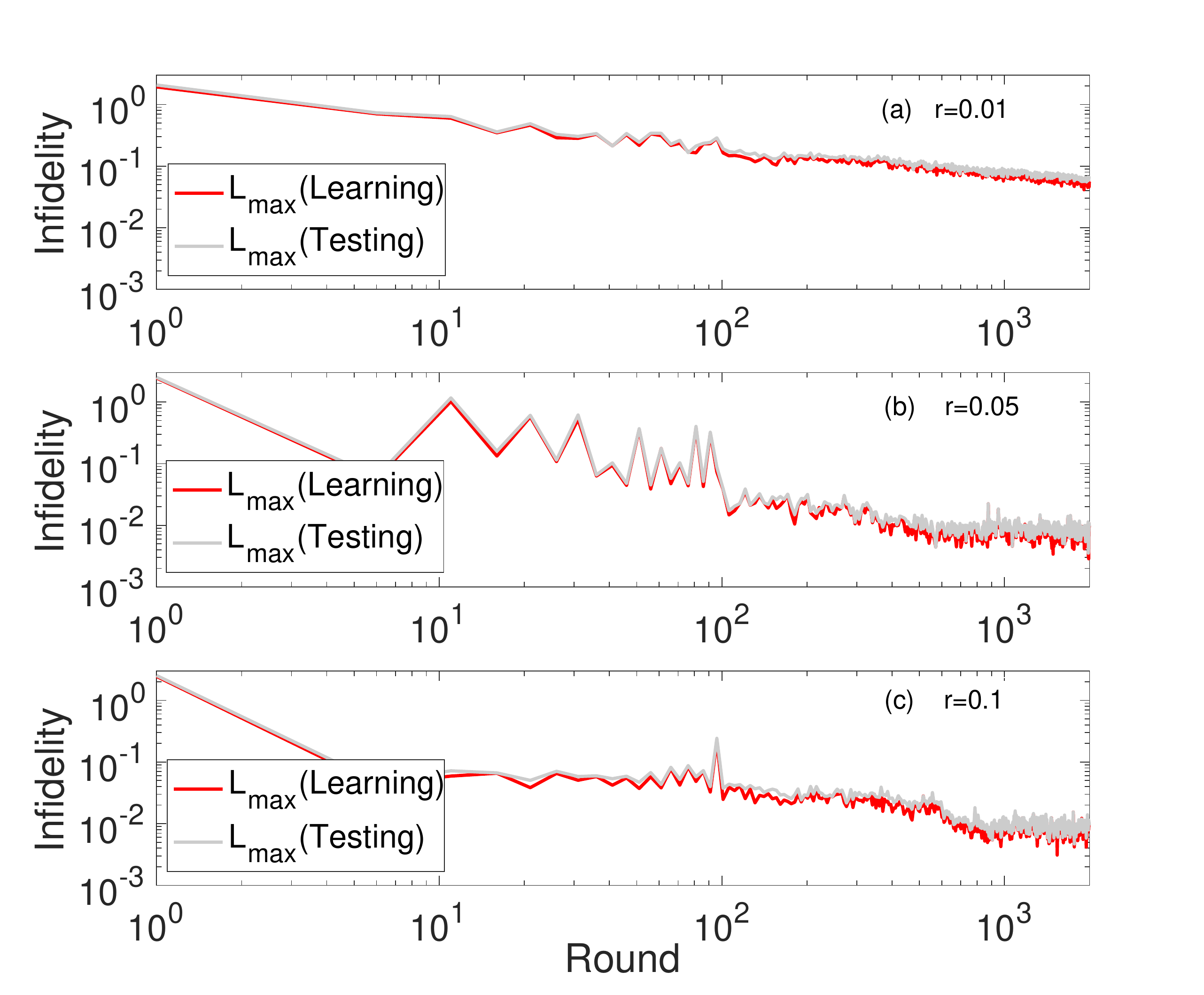}
 	\end{center}
 	\caption{ The worst-case infidelity versus the number of rounds in better-response a-GRAPE optimization of the three-qubit system with $r=0.01,\ 0.05$ and $0.1$.
 	}\label{betterlearning}
 \end{figure}

For the better-response approach, we choose $M=100$ and compare the performance under $r=0.01,\ 0.05,$ and $0.1$, where the true worst-case infidelity in each round is estimated by 2000 independent random samples. Similar to the case of the best-response approach, the simulation results (see the learning curves in Fig.~\ref{betterlearning}) show that the robustness of the optimized controls can be improved by using adequately many adversarial samples, but too many will not bring further improvement.

\begin{figure}
	\begin{center}
		\includegraphics[width=1\columnwidth]{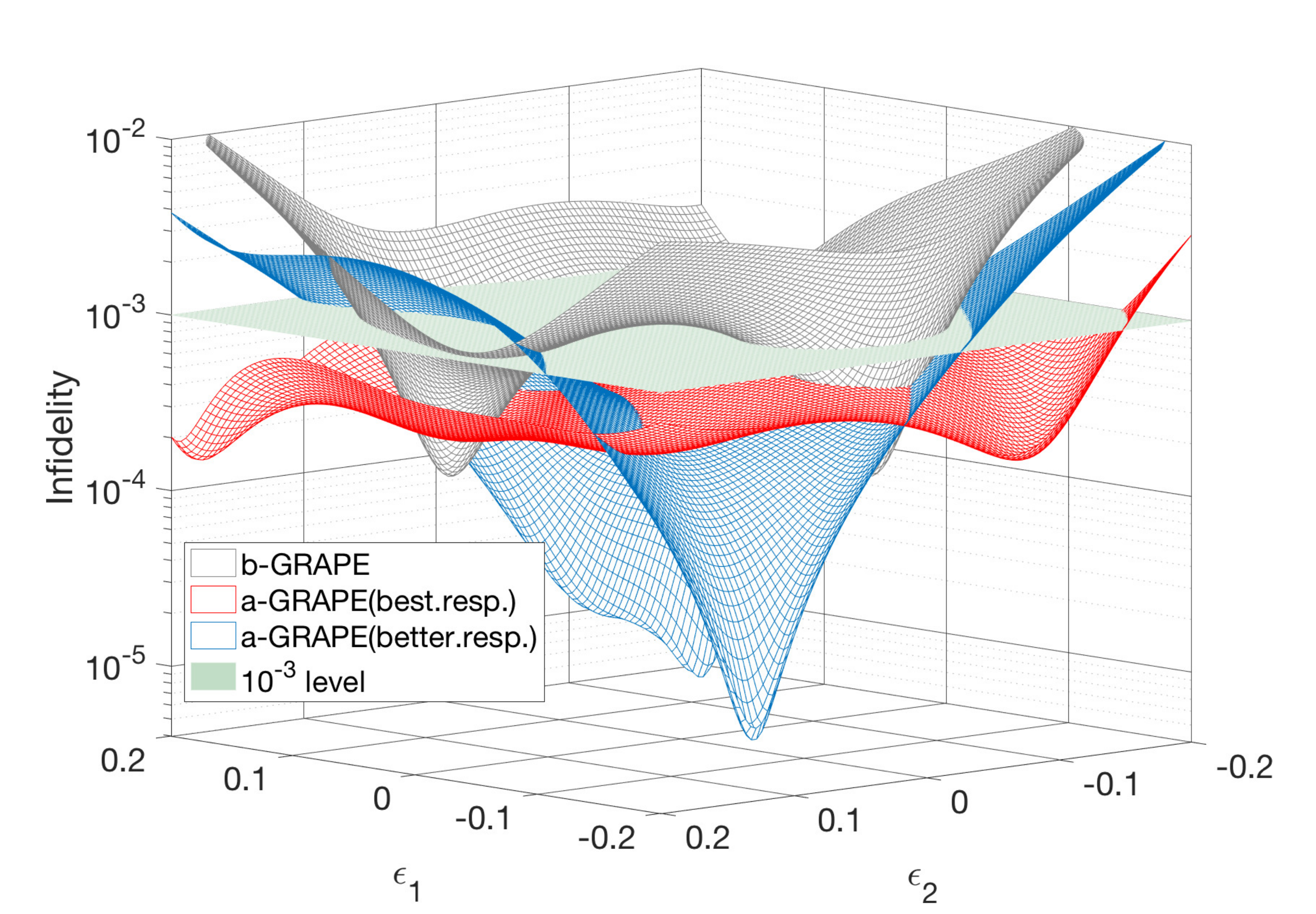}
	\end{center}
	\caption{ The infidelity versus uncertainty parameters under controls optimized with b-GRAPE with $n_{mb}=1$ and $\alpha=0.002$, best-response dynamic with memory size $s=10$, better-response approach with ratio $r=0.05$ respectively.
	}\label{3D}
\end{figure}

Since there are only two uncertainty parameters in this example, we can plot a 3D landscape to show how the infidelity varies with them, from which we can evaluate the overall robustness. In Fig.~\ref{3D}, we display 3D plots under controls optimized with the algorithms b-GRAPE (after 2 million iterations), best-response a-GRAPE (after 629 rounds) and better-response a-GRAPE (after 1986 rounds). The comparison shows that both a-GRAPE algorithms outperform the b-GRAPE algorithm, as most of their landscape surfaces are below that of b-GRAPE. The best-response a-GRAPE achieves the lowest worst-case infidelity and effectively suppresses almost the entire landscape down below the level of $L=10^{-3}$. However, its overall performance in the higher-precision regime (e.g., $L=10^{-4}$) is poorer than the better-response a-GRAPE.

In addition to the best-response and better-response approaches, the a-GRAPE algorithms can be designed more flexibly such that it still works efficiently when the uncertainty vector $\boldsymbol{\epsilon}$ is of large scale. For example, we may also perform the minimization process in a relaxed manner. As is described in Algorithm~\ref{algorithm3}, we may update the control by only a few gradient-descent iterations (or stop at some prescribed error threshold) without having to reach the ultimate minimum, which responds better but not best to the adversarial samples. The maximization part can be done either with the best-response or better-response approaches. In this regard, the SCP algorithm~\cite{kosut2013robust} can be considered as a special case of the relaxed better-response approach with fixed sampled uncertainties and carefully selected learning rates. It is noteworthy that as the minimization and maximization are related to the fidelity and the robustness respectively, the balance between the two performance indexes can be adjusted flexibly via relaxing the two optimization processes.

\begin{figure}
	\begin{center}
		\includegraphics[width=1\columnwidth]{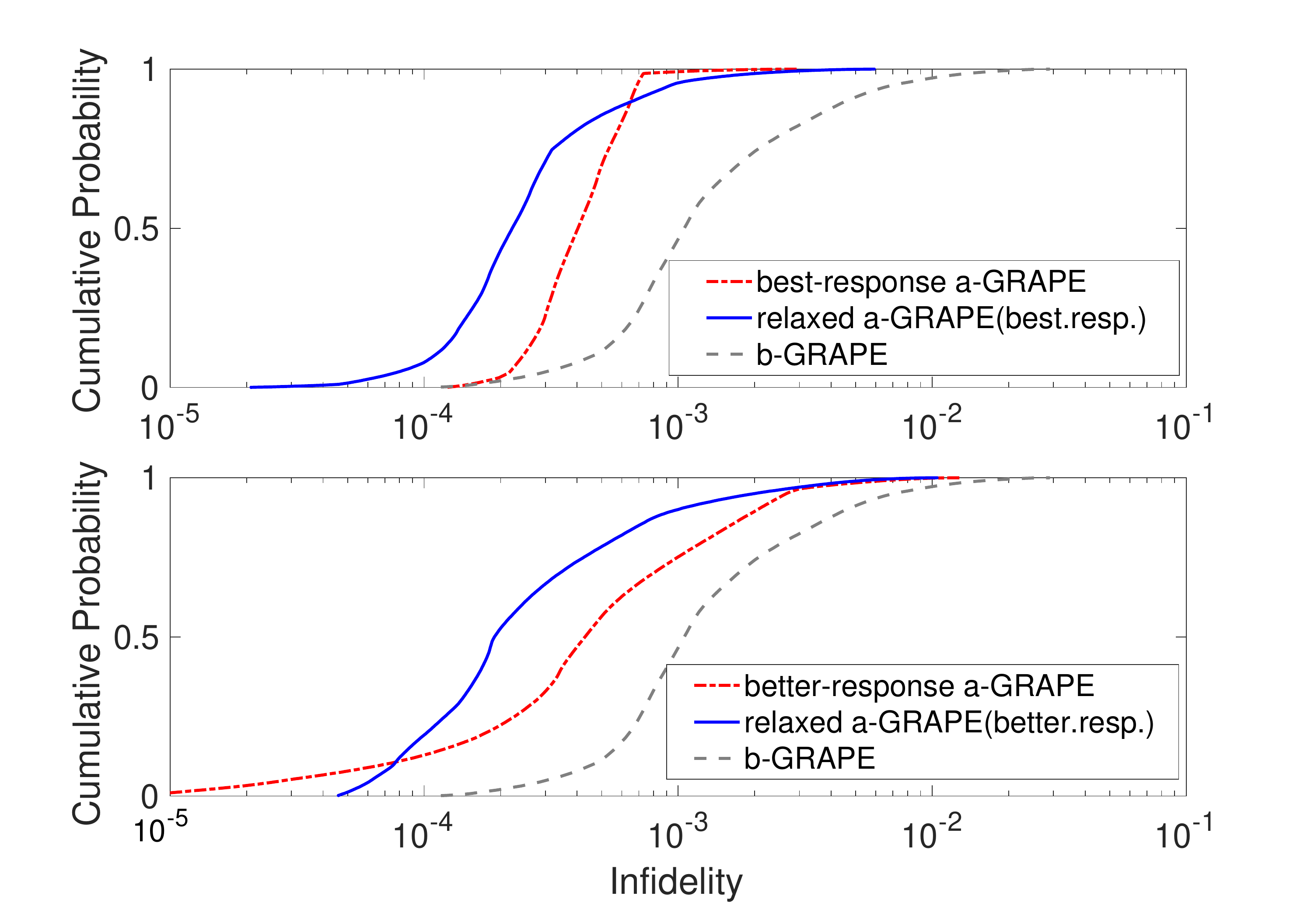}
	\end{center}
	\caption{The cumulative probability versus infidelity under the corresponding controls used in Fig.~\ref{3D} and the ones obtained from the relaxed best-response approach run by 10995 rounds and the relaxed better-response algorithm run by 3850 rounds.}
	\label{3bitcdf}
\end{figure}

To assess the feasibility of this idea, we apply the relaxed best-response (with $s=5$, $n=20$ and $m=20$) and relaxed better-response (with $r=0.25$, $n=30$ and $m=20$) a-GRAPE algorithms to the same three-qubit example. As is shown in Fig.~\ref{3bitcdf}, where the control robustness is evaluated by the cumulative probability functions, the relaxed a-GRAPE algorithms can also greatly outperform the b-GRAPE algorithm. Compared to their unrelaxed counterparts, the relaxed best-response and relaxed better-response a-GRAPE algorithms are not only faster, but also more robust in the high-precision regime (e.g., near the infidelity level $10^{-4}$). 

\begin{algorithm}[htb]\label{algorithm3}
\caption{relaxed \textit {best/better-response} a-GRAPE}

\noindent{\bfseries Initialize:}

 \setlength\parindent{1em} a randomly chosen initial control $\boldsymbol{u}$;	
 
  \setlength\parindent{1em} a set memory size $s$ or ratio $r$;
  
  \setlength\parindent{1em} an initial adversarial sample set $B$.

\noindent{\bfseries Repeat:}
\begin{itemize}[itemsep=1pt, topsep=2pt,partopsep=2pt,parsep=1pt]
	\item[(1)] Randomly generate $m$ uncertainty samples, and select the worst one or the first $rm$ worst ones;
	\item[(2)] Do the following GRAPE optimization for $n$ iterations:
	{\setlength\abovedisplayskip{1pt}
		\setlength\belowdisplayskip{1pt}
			\begin{equation*} \boldsymbol{u}\leftarrow\boldsymbol{u}-\alpha\cdot \frac{\delta}{\delta \boldsymbol{u}}J[\boldsymbol{u},B].
		\end{equation*}}
Here, $\alpha$ represents the learning rate;
\item[(3)] Update the adversarial sample set $B$ as described in Algorithm 1 or 2;
	\end{itemize}
	
	\noindent {\bfseries End} if the stoping criteria are satisfied.
	\end{algorithm}

\section{Conclusion}\label{conclusion}

We have proposed a family of adversarial learning algorithms, including the best-response and better-response approaches, for the robust control design of quantum systems. The algorithms are subject to the optimization of worst-case gate infidelity, which can be treated and resolved from a game-theoretic perspective. Numerical simulations show that these a-GRAPE algorithms can achieve high control robustness. In particular, the best-response approach can effectively suppress the error over a larger domain at a satisfactory level of precision, but in the extremely high-precision regime, the better-response approach is superior. Both approaches have their regimes of practical utility depending on the application-specific requirement of the control. We also demonstrate that a family of a-GRAPE algorithms can be expanded by relaxing the maximization and minimization processes.

It should be noted that, although a-GRAPE usually outperforms b-GRAPE, the computational burden is also heavier, and the tuning of algorithm parameters (e.g., memory sizes, batch sizes, ratios or learning rates) is application-specific. Further studies are needed to deduce how to optimize the choices of these parameters, or even in an adaptive fashion. 

 As we remarked, we did not require on the existence of Nash equilibria (NE) as an appropriate adversarial algorithm can enhance the robustness no matter whether the NE exists or not. From our simulations, it appears that a mixed NE is more likely approached in the best-response a-GRAPE with a larger memory size. From a theoretical perspective, a better understanding of the existence of a NE will be useful to attain. This topic will be explored in future studies.

\begin{acknowledgements}
The author Rebing Wu acknowledges the support of the National Key R$\&$D Program of China (Grants No. 2018YFA0306703 and No. 2017YFA0304304) and NSFC (Grants No. 61833010 and No. 61773232). The author Herschel Rabitz acknowleges the support of the National Science Foundation (CHE-1763198).
\end{acknowledgements}

\appendix*
\section{b-GRAPE algorithm}
The b-GRAPE algorithm presented in~\cite{Wu2018Deep} is a stochastic gradient algorithm. The optimization process follows the gradient evaluated with randomly chosen batches of samples, so that the uncertainties can be effectively used to improve the robustness. Here, ``b" stands the ``batch". The b-GRAPE algorithm is described in Algorithm~\ref{bgrape}.

\begin{algorithm}[htb]\label{bgrape}
	\caption{ b-GRAPE}
	
	\noindent{\bfseries Initialize:}
	
	\setlength\parindent{1em} a randomly choosen initial control $\boldsymbol{u}^{(0)}$;	
	
	\setlength\parindent{1em} an initial momentum $\boldsymbol{v}^{(0)}=\boldsymbol{0}$;	
	
	\setlength\parindent{1em} a set mini-batch size $n_{mb}$ ;
	
	\noindent{\bfseries Repeat:}
	\begin{itemize}[itemsep=1pt, topsep=2pt,partopsep=2pt,parsep=1pt]
		\item[(1)] Randomly select a subset $\mathcal{S}_k$ of the uncertainty sample with $|\mathcal{S}|_k=n_{mb}$, where $k$ denotes the current number of round;
		\item[(2)] Update the control by
		{\setlength\abovedisplayskip{1pt}
			\setlength\belowdisplayskip{1pt}
			\begin{equation*} \boldsymbol{u}^{(k)}=\boldsymbol{u}^{(k-1)}+\boldsymbol{v}^{(k)},
			\end{equation*}
			with
			\begin{equation*}
			\boldsymbol{v}^{(k)}=\lambda\boldsymbol{v}^{(k-1)}-\frac{\alpha}{n_{mb}}\sum_{\boldsymbol{\epsilon}\in \mathcal{S}_k}\frac{\delta}{\delta \boldsymbol{u}}L[\boldsymbol{ u}^{(k-1)},\boldsymbol{\epsilon}].
			\end{equation*}}
		Here, $\alpha$ represents the learning rate and the weight parameter $\lambda$ is chosen to be $0.9$ in this paper;
	\end{itemize}
	
	\noindent {\bfseries End} if the stoping criteria are satisfied.
\end{algorithm}

\bibliographystyle{unsrt}
\bibliography{gamelearning}

\end{document}